\documentclass[11pt]{article}

\usepackage[left=2cm, right=2cm, top=2.5cm, bottom=2.5cm]{geometry}
\newsavebox{\foobox}
\newcommand{\slantbox}[2][0]{\mbox{%
        \sbox{\foobox}{#2}%
        \hskip\wd\foobox
        \pdfsave
        \pdfsetmatrix{1 0 #1 1}%
        \llap{\usebox{\foobox}}%
        \pdfrestore
}}
\newcommand\unslant[2][-.25]{\slantbox[#1]{$#2$}}

\newcommand{\mpi}{\text{\unslant[-.18]\pi}}
\newcommand{\mdelta}{\text{\unslant[-.15]\delta}}

\usepackage{isomath}
\usepackage[x11names]{xcolor}
\usepackage{fancyhdr, amssymb, cancel, amsmath, graphicx, pgfplots, tikz}

\newcommand{\stylecolor}{black}

\usepackage[colorlinks=true, urlcolor=violet, linkcolor=blue, citecolor=red, hyperindex=true, linktocpage=true]{hyperref}

\usepackage[explicit]{titlesec}

\newcommand*\sectionlabel{}
\titleformat{\section}
  {\gdef\sectionlabel{}
   \Large\bfseries\scshape}
  {\gdef\sectionlabel{\thesection. }}{0pt}
  {\begin{tikzpicture}[remember picture,overlay]
	\draw (-0.2, 0) node[right] {\textsf{\sectionlabel#1}};
	\draw[thick] (0, -0.4) -- (\textwidth, -0.4);
       \end{tikzpicture}
  }
\titlespacing*{\section}{0pt}{15pt}{20pt}

\newcommand*\subsectionlabel{}
\titleformat{\subsection}
  {\gdef\subsectionlabel{}
   \large\bfseries\scshape}
  {\gdef\subsectionlabel{\thesubsection.\ \  }}{0pt}
  {\begin{tikzpicture}[remember picture,overlay]
    	\draw (-0.15, 0) node[right] {\textsf{\subsectionlabel#1}};
       \end{tikzpicture}
  }
\titlespacing*{\subsection}{0pt}{10pt}{10pt}

\newcommand*\subsubsectionlabel{}
\titleformat{\subsubsection}
  {\gdef\subsubsectionlabel{}
   \bfseries\scshape}
  {\gdef\subsubsectionlabel{\thesubsubsection.\ \  }}{0pt}
  {\begin{tikzpicture}[remember picture,overlay]
    	\draw (-0.15, 0) node[right] {\textsf{\subsubsectionlabel#1}};
       \end{tikzpicture}
  }
\titlespacing*{\subsubsection}{0pt}{7pt}{7pt}

\pgfplotsset{every axis legend/.append style={at={(1.02,1)},anchor=north west}}

\newcommand{\titletext}{Conductivity of a strange metal: \\  from holography to memory functions}

\begin{document}

\allowdisplaybreaks

\pagestyle{fancy}
\renewcommand{\headrulewidth}{0pt}
\fancyhead{}

\fancyfoot{}
\fancyfoot[C] {\textsf{\textbf{\thepage}}}

\begin{equation*}
\begin{tikzpicture}
\draw (0.5\textwidth, -3) node[text width = \textwidth] {{\huge \begin{center} \color{\stylecolor} \textsf{\textbf{\titletext}} \end{center}}};
\end{tikzpicture}
\end{equation*}
\begin{equation*}
\begin{tikzpicture}
\draw (0.5\textwidth, 0.1) node[text width=\textwidth] {\large \color{black} $\text{\textsf{Andrew Lucas}}$};
\draw (0.5\textwidth, -0.5) node[text width=\textwidth] {\small \textsf{Department of Physics, Harvard University, Cambridge, MA 02138, USA}};
\end{tikzpicture}
\end{equation*}
\begin{equation*}
\begin{tikzpicture}
\draw (0.5\textwidth, -6) node[below, text width=0.8\textwidth] {\small  We study the electrical response of a wide class of strange metal phases without quasiparticles at finite temperature and charge density, with explicitly broken translational symmetry, using holography.    The low frequency electrical conductivity exhibits a Drude peak, so long as momentum relaxation is slow.   The relaxation time and the direct current conductivity are exactly equal to what is computed, independently of holography, via the memory function framework.};
\end{tikzpicture}
\end{equation*}
\begin{equation*}
\begin{tikzpicture}
\draw (0, -13.1) node[right, text width=0.5\paperwidth] {\texttt{lucas@fas.harvard.edu}};
\draw (\textwidth, -13.1) node[left] {\textsf{\today}};
\end{tikzpicture}
\end{equation*}

\tableofcontents

\section{Introduction}
Holographic duality has provided condensed matter physicists with a novel way to study the properties of strongly correlated transport and dynamics in theories without quasiparticles \cite{review1, review2, review3}.   Charged black holes become dual to states of matter at finite charge density and temperature.    At finite density, however, it becomes crucial to include the effects of momentum relaxation in order to obtain a finite direct electrical conductivity, $\sigma_{\mathrm{dc}}$, at zero frequency and momentum.\footnote{This follows directly from the fact that at finite density, one can generate an electrical current without any energy in a translationally invariant theory by simply boosting to a reference frame with a relative velocity to the rest frame of the metal.}  These momentum relaxing effects may either be studied by numerically constructing a charged black hole which breaks translational invariance \cite{santos, ling, donos1409}, or analytically, so long as the black hole breaks translation invariance only weakly \cite{hartnollhofman, blake2, lss, LS2}, translational invariance is broken via massive gravity \cite{vegh, davison, blake1, dsz}, ``helical lattices" \cite{hartnolldonos, donos1406} or  ``Q-lattices" \cite{donos1, andrade, gouteraux}.  

In this paper we show, through a direct holographic computation, that the low-frequency conductivity  is given by a Drude peak in a wide class of holographic metals where translational symmetry is weakly broken: \begin{equation}
\sigma(\omega) =  \frac{\sigma_{\mathrm{dc}}}{1-\mathrm{i}\omega \tau}. \label{drude}
\end{equation}These metals have coherent transport without quasiparticles, in the language of \cite{hartnollscale1}.  Furthermore, $\sigma_{\mathrm{dc}}$ and $\tau$ are identical to the results computed -- independently of holography -- through the memory function formalism \cite{zwanzig, mori, forster}.  Memory functions have been used, in conjunction with holographic methods, to evaluate conductivities in the past \cite{hartnollhofman, hartnollimpure}, as well as in non-holographic condensed matter models for strange metals as well \cite{rosch, hkms, raghu, patel}.    More recently it has been noted that exact computations of $\sigma_{\mathrm{dc}}$ -- the direct current conductivity at zero momentum and frequency -- admit temperature scaling in agreement with results found through memory functions \cite{blake2, lss}.   In fact, the correspondence hinted at in these works is exact, including all other prefactors.     A short computation within the memory function framework necessitates the emergence of a Drude peak (though with complications when magnetic fields and/or strong charge diffusion is present \cite{LS}).   It is not obvious from a purely holographic computation why this Drude peak should be universal, and how it should arise -- other than resorting to the argument that these metals should be describable by the memory function formalism.   We will address these points in this paper.

We begin with a review of why the Drude peak is universal, via hydrodynamic arguments, in Section \ref{sec2}.  We then describe how to the Drude peak arises within the memory function formalism in Section \ref{sec3}.   Section \ref{sec4} describes the holographic computation of $\sigma(\omega)$, and a proof of equivalence with the results of Section \ref{sec3}.

\section{Hydrodynamics and the Drude Peak} \label{sec2}
Drude peaks generically emerge from quantum field theories described by hydrodynamics with a small amount of momentum relaxation.    Unlike the historical model of Drude physics (electrons scattering off of a lattice or impurities), quasiparticles need not exist to observe (\ref{drude}).  Consider the long-time equation for momentum relaxation in a charged fluid, placed in an electric field in the $x$-direction: \begin{equation}
\partial_t \Pi_x + \partial_x P = -\frac{\Pi_x}{\tau} + \mathcal{Q} E_x. 
\end{equation}Here $\Pi_x$ is the $x$-momentum density, $P$ is the pressure, $\tau$ is the momentum relaxation time, and $\mathcal{Q}$ is the charge density (at rest) of the fluid.    As we are interested in zero-momentum transport, at frequency $\omega$ this equation simplifies to \begin{equation}
\left(\frac{1}{\tau} -\mathrm{i}\omega\right)\Pi_x = \mathcal{Q}E_x.
\end{equation}To compute \begin{equation}
\sigma_{\mathrm{dc}} = \frac{J_x}{E_x},
\end{equation}we need to relate $J_x$ to $\Pi_x$.

At this point, let us specialize further to a relativistic quantum field theory, and work in units where $\hbar=c=1$.\footnote{Note that, from the perspective of condensed matter physics, ``$c$" here is the effective emergent speed of light in the low energy Lorentz-invariant theory.}  All holographic models we will study will fall into this class of theories.   In such a theory, we find that at small velocities $v_x$, the momentum density is given by \begin{equation}
\Pi_x = (\epsilon+P)v_x,  \label{pix}
\end{equation}where $\epsilon$ is the energy density of the fluid at rest and $P$ is the pressure, and \begin{equation}
J_x = \mathcal{Q}v_x. \label{jx}
\end{equation}
Combining these equations we can relate $J_x$ to $\Pi_x$, and find a Drude conductivity of the form (\ref{drude}) with \begin{equation}
\sigma_{\mathrm{dc}} = \frac{\mathcal{Q}^2\tau}{\epsilon+P}.  \label{sigmadc}
\end{equation}
This is a special limiting case of more general hydrodynamic results found in \cite{hkms}.  It is valid when charge diffusion is negligible, and there is no magnetic field.     

\section{Memory Functions} \label{sec3} 
Hydrodynamics does not give us an explicit expression for any thermodynamic quantities, or $\tau$.   The thermodynamic functions such as $\epsilon$ and $P$ are ``intrinsic" to the quantum field theory, and are approximately independent to the precise mechanism of momentum relaxation (so long as $\tau^{-1}$ is small).   On the other hand, $\tau$ is ``extrinsic" and sensitive to the precise way in which momentum can relax.   The memory function analysis provides us with a way of computing $\tau$, given that we know the microscopic method by which momentum can relax.

More precisely, in a time-reversal symmetric theory in which momentum is the only almost conserved quantity, the memory matrix framework tells us that \begin{equation}
\sigma(\omega) = \frac{\chi_{JP}^2}{M_{PP}(\omega)-\mathrm{i}\omega \chi_{PP}}, \label{sigmamm}
\end{equation}where $\chi_{PP}$ is the momentum-momentum susceptibility, $\chi_{JP}$ is the current-momentum susceptibility, and $M_{PP}(\omega)$ is the momentum-momentum component of the memory matrix.    If there are other long-lived conserved quantities, then a matrix generalization of this equation applies \cite{forster}.   For us, $M_{PP}(\omega)$ is a small quantity and may thus be computed perturbatively.   In our relativistic field theories, one finds \begin{subequations}\label{chieq}\begin{align}
\chi_{PP} &= \epsilon+P, \\
\chi_{JP} &= \mathcal{Q}.
\end{align}\end{subequations}To derive these results, we note that the conjugate thermodynamic variable to momentum $P_x$ is velocity $v_x$.   The susceptibility $\chi_{\alpha P_x}$ is equal to $\langle \alpha\rangle/v_x$ as $v_x\rightarrow 0$.  (\ref{chieq}) then follows from (\ref{pix}) and (\ref{jx}).     As $\omega\rightarrow 0$, one finds \cite{hartnollhofman} \begin{equation}
M_{PP}(0) = \lim_{\omega\rightarrow 0} \frac{\mathrm{Im}\left(G^{\mathrm{R}}_{\dot{P}_x\dot{P}_x}(\omega)\right)}{\omega} \equiv \frac{\epsilon +P}{\tau},
\end{equation}where $\dot{P}_x$ is the time derivative of the total $x$-momentum.   We will argue in the course of our holographic discussion that finite frequency corrections to the memory matrix will be extremely small in the regime we are interested in.   $\tau$ may thus be computed from this memory matrix component at $\omega=0$, and a simple calculation verifies that (\ref{drude}) and (\ref{sigmadc}) are recovered.

Now, let us specialize further to a quantum field theory with a translationally invariant Hamiltonian $H_0$, perturbed to \begin{equation}
H = H_0 - \int \mathrm{d}^d\mathbf{x} \; h(\mathbf{x}) \mathcal{O}(\mathbf{x}),  \label{ham}
\end{equation}where $\mathcal{O}$ is a Lorentz scalar operator in the original theory, and $h(\mathbf{x})$ is a (real-valued) static, spatially varying field.  The quantum operator \begin{equation}
\dot{P}_x  = \mathrm{i}[P_x, H] = -\mathrm{i}\int\mathrm{d}^d\mathbf{x} \; h(\mathbf{x}) [P_x,\mathcal{O}(\mathbf{x})] =  \int \mathrm{d}^d\mathbf{x}  \; h(\mathbf{x}) (\partial_x\mathcal{O})(\mathbf{x})
\end{equation}
  We then employ \begin{equation}
G^{\mathrm{R}}_{\dot{P}_x\dot{P}_x}(\omega) = \int \mathrm{d}^d\mathbf{k}\mathrm{d}^d\mathbf{q} \;  h(\mathbf{k})\overline{h(\mathbf{q})}  G^{\mathrm{R}}_{\partial_x \mathcal{O} \partial_x\mathcal{O}}(\mathbf{k},\mathbf{q},\omega) 
 \end{equation}The perturbative parameter that makes $M_{PP}$ small is $\tau^{-1}\sim h^2$.  At leading order in $h$, $G^{\mathrm{R}}$ is the Green's function of a translationally invariant theory and so the integral identically vanishes when $\mathbf{k}+ \mathbf{q} \ne \mathbf{0}$.    
 
If we have a ``lattice"  or periodic potential where \begin{equation}
 h(\mathbf{x}) = h_0 \cos(\mathbf{k}_0\cdot\mathbf{x}),
 \end{equation}
 then we obtain \begin{equation}
 G^{\mathrm{R}}_{\dot{P}_x\dot{P}_x}(\omega) = \frac{h_0^2}{2}k_0^2 G^{\mathrm{R}}_{\mathcal{O}\mathcal{O}}(\mathbf{k}_0,\omega).
 \end{equation}In the case of disorder, the function $h(\mathbf{x})$ will be random.   It is common to take $h(\mathbf{x})$ to be a zero-mean Gaussian random function with mean and variance given by ($\mathbb{E}[\cdots]$ denotes disorder averaging) \begin{subequations}\begin{align}
\mathbb{E}[ h(\mathbf{x})] &= 0, \\
\mathbb{E}[h(\mathbf{x})h(\mathbf{y})] &= \varepsilon^2 \mdelta(\mathbf{x}-\mathbf{y}).   \label{epsvar}
\end{align}\end{subequations}We obtain \cite{hartnollimpure}\begin{align}
 G^{\mathrm{R}}_{\dot{P}_x\dot{P}_x}(\omega) &= \int \mathrm{d}^d\mathbf{k}\mathrm{d}^d\mathbf{q}\; \mathbb{E}[h(\mathbf{k})h(\mathbf{q})]  k_x^2 G^{\mathrm{R}}_{\mathcal{OO}}(\mathbf{k},\mathbf{q},\omega)  = \int \mathrm{d}^d\mathbf{k}\mathrm{d}^d\mathbf{q}\; \varepsilon^2  k_x^2 \mdelta(\mathbf{k}+\mathbf{q}) G^{\mathrm{R}}_{\mathcal{OO}}(\mathbf{k},\mathbf{q},\omega) \notag \\
 &=  \int \mathrm{d}^d\mathbf{k} \; \varepsilon^2 k_x^2 G^{\mathrm{R}}_{\mathcal{O}\mathcal{O}}(\mathbf{k},\omega).  \label{eq17}
\end{align}
We approximate $G^{\mathrm{R}}_{\dot{P}_x\dot{P}_x}$ by its average as fluctuations are suppressed in the large volume limit  \cite{LS2}.   More generally, if we assume that $h(\mathbf{x})$ and its derivatives are non-vanishing (almost) everywhere in space, we will find \begin{equation}
 G^{\mathrm{R}}_{\dot{P}_x\dot{P}_x}(\omega) \; ``="  \int \mathrm{d}^d\mathbf{k} |h(\mathbf{k})|^2 k_x^2 G^{\mathrm{R}}_{\mathcal{O}\mathcal{O}}(\mathbf{k},\omega).  \label{eq18}
\end{equation}
We have belabored this rather trivial discussion to emphasize the following -- in the last equation, the quotes around the equals sign arise because technically, we have neglected a $\mdelta$ function enforcing momentum conservation in $|h(\mathbf{k})|^2$.\footnote{For example, in the disorder case, using $\overline{h(-\mathbf{k})} = h(\mathbf{k})$ and $\mathbb{E}[h(\mathbf{k})h(\mathbf{q})] = \varepsilon^2 \mdelta(\mathbf{k}+\mathbf{q})$, (but without the $\mdelta$ function) in (\ref{eq18}), we obtain (\ref{eq17}).   Crudely, one should think of dividing $|h(\mathbf{k})|^2$ by $\mdelta(\mathbf{0})$ in (\ref{eq18}).}  However, writing the equation in the form (\ref{eq18}) is convenient: we will see a similar equation with a ``missing" $\mdelta$ function arise holographically.     Thus \begin{equation}
\frac{\epsilon+P}{\tau} \equiv \int \mathrm{d}^d\mathbf{k} \; |h(\mathbf{k})|^2 k_x^2 \times \lim_{\omega\rightarrow 0} \frac{\mathrm{Im}\left( G^{\mathrm{R}}_{\mathcal{O}\mathcal{O}}(\mathbf{k},\omega)\right)}{\omega}.  \label{taudef} 
\end{equation}  

\section{Holography} \label{sec4}
Now let us turn to holography.   We assume that $d>1$.  We consider solutions of the Einstein-Maxwell-dilaton (EMD) system with action \begin{equation}
S = \int \mathrm{d}^{d+2}x \sqrt{-g}\left(\frac{1}{2\kappa^2}\left(R-2(\partial_M\Phi)^2 - \frac{V(\Phi)}{L^2}\right) - \frac{Z(\Phi)}{4e^2}F_{RS}F^{RS} \right),
\end{equation}where $g_{MN}$ is the bulk metric (dual to the stress tensor of the boundary theory), $A_M$ is a U(1) gauge field (dual to the electric current of the boundary theory), and $\Phi$ is a dilaton field.     These holographic models are known to give rise to rich families of boundary theories; we focus on metallic phases in this paper.

Let us now, without proof, state some results about static, isotropic geometries that solve the equations of motion associated with this EMD action.    We use the conventions and results of \cite{LS2} in what follows.     These solutions have metric \begin{equation}
\mathrm{d}s^2 = \frac{L^2}{r^2}\left[\frac{a(r)}{b(r)} \mathrm{d}r^2 - a(r)b(r)\mathrm{d}t^2 + \mathrm{d}\mathbf{x}^2\right],
\end{equation}where $b(r)$ plays the role of an ``emblackening factor":   in particular, near the black hole horizon (of planar topology), located at finite $r=r_{\mathrm{h}}$, we find \begin{equation}
b(r)\approx 4\mpi T(r_{\mathrm{h}}-r),  \label{beq}
\end{equation}where $T$ is the Hawking temperature of the black hole, which also equals the dual field theory's temperature;  $a$ is finite near the horizon.   Near the boundary ($r=0$),  $a(0)=b(0)=1$;  $\Phi(r=0)=0$, $Z(0)=1$ and $V(0)=-d(d+1)$;  the asymptotic geometry is that of AdS.   The UV of the continuum field theory is thus approximately conformal.   The profile of the gauge field is \begin{equation}
A = p(r)\mathrm{d}t
\end{equation}where $p(r_{\mathrm{h}})=0$, and the asymptotic behavior near the boundary is \begin{equation}
p(r)\approx \mu - \frac{r^{d-1}}{d-1}\frac{e^2\mathcal{Q}}{L^{d-2}} + \cdots.   \label{peq2}
\end{equation}$\mu$ is the chemical potential associated with the conserved charge.  One finds that the object $\mathcal{C}$, defined as\begin{equation}
\mathcal{C} \equiv \frac{(ab)^\prime}{ar^d} + \frac{2\mathcal{Q}\kappa^2}{L^d}p  \label{peq}
\end{equation}is independent of $r$, where here and henceforth, primes denote $r$-derivatives. There is, of course, much more that can be said about these geometries, but this is all we will need for the present paper.  Although we have assumed (for technical ease) that the UV geometry is AdS,  the remainder of the geometry may be completely arbitrary, so long as it may be constructed as a solution of EMD theory.

Now, we will add to this geometry a fourth field:  a neutral scalar $\psi$, dual to the operator $\mathcal{O}$ sourced by the translation symmetry breaking field $h(\mathbf{x})$.   The action of $\psi$ is \begin{equation}
S_\psi = -\frac{1}{2} \int \mathrm{d}^{d+2}x\sqrt{-g} \left( (\partial_M\psi)^2 + B(\Phi)\psi^2\right),  \label{spsi}
\end{equation}
If the operator $\mathcal{O}$ has (UV) dimension $\Delta > (d+1)/2$ (we choose this so that the operator $\mathcal{O}$ is described by standard quantization in holography), then the near-boundary asymptotic expansion of $\mathcal{O}$ is \begin{equation}
\psi(r) = r^{d+1-\Delta} \frac{\psi^{(0)}(\mathbf{x})}{L^{d/2}}  + \cdots + r^\Delta \frac{\psi^{(1)}(\mathbf{x})}{L^{d/2}} + \cdots
\end{equation}and $B(0)= \Delta(\Delta-d-1)/L^2$.      In general, $B$ should be chosen non-trivially in order to obtain results which match boundary theory expectations \cite{lss}.
The boundary conditions which imply that the Hamiltonian of the boundary theory is given by (\ref{ham}) are imposed by setting \begin{equation}
\psi^{(0)}(\mathbf{x}) = h(\mathbf{x}).   \label{psi0h0}
\end{equation}
This will induce a small backreaction on the original EMD system, as the bulk stress tensor of $\psi$ is a source in Einstein's equations.   However, this source is $\mathrm{O}(h^2)$, and following \cite{blake2, lss} we may treat the geometry at $\mathrm{O}(h)$ (thus the background metric is unperturbed) in order to compute the conductivity at leading order in $h$ (as will be clear in the derivation shortly).    We may thus take \begin{equation}
\psi = \int \mathrm{d}^d\mathbf{k} \; h(\mathbf{k}) \psi_0(\mathbf{k},r) \mathrm{e}^{\mathrm{i}\mathbf{k}\cdot\mathbf{x}},
\end{equation}with $\psi_0(\mathbf{k},r\rightarrow 0)\sim L^{-d/2}r^{d+1-\Delta}$, and $\psi_0$ obeying the linear equation of motion associated with the action (\ref{spsi}), subject to a regularity condition at $r=r_{\mathrm{h}}$, and the boundary condition (\ref{psi0h0}).  A holographic calculation can explicitly determine the scale of $h$ at which this approximation breaks down \cite{lss}.     

The expectation value of $\mathcal{O}(\mathbf{x})$ in our state at finite density and temperature, and in the background field $h$, is given by \cite{skenderis} \begin{equation}
\langle \mathcal{O}(\mathbf{x})\rangle  = (2\Delta-d-1) \psi^{(1)}(\mathbf{x}).   \label{expo}
\end{equation}
If the asymptotic expansion of $\delta A_x$ near $r=0$ is \begin{equation}
\delta A_x = \delta A_x^{(0)}  + \frac{r^{d-1}}{d-1} \delta A_x^{(1)} + \cdots,
\end{equation}then \begin{equation}
\sigma(\omega) = \frac{L^{d-2}}{\mathrm{i}\omega e^2} \frac{\delta A_x^{(1)}}{\delta A_x^{(0)}}.  \label{sigmaomega}
\end{equation}
We will use these facts to compute Green's functions and conductivities in this section.

\subsection{Drude Peak}
To compute the conductivity, we need to compute the response of our background EMD-scalar solution to a perturbation $\delta A_x \mathrm{e}^{-\mathrm{i}\omega t}$ (at zero momentum) -- equivalent in the boundary theory to imposing a small electric field.   The solution need only be found within linear response theory, as is standard, but the computation below will proceed a bit different than that in \cite{blake2, lss}, as we must explicitly consider finite $\omega$ effects.      $\delta A_x$ can only consider spin 1 perturbations under the spatial $\mathrm{SO}(d)$ isometry of the metric.   These perturbations are (in axial gauge $\delta g_{rx}=0$): \begin{equation*}
\delta A_x, \;  \delta\tilde{g}_{tx} \equiv \frac{r^2}{L^2}\delta g_{tx}, \; \partial_x \delta \psi, \; \partial^2 \partial_x \delta \psi, \ldots
\end{equation*}
By computing the full linear response problem (we discuss the remainder of the necessary boundary conditions later), we can find the near boundary asymptotic behavior of $\delta A_x(r)$ and compute $\sigma(\omega)$ from (\ref{sigmaomega}).

Linearizing the EMD system, one finds the equations of motion: \begin{subequations}\label{emdlin}\begin{align}
\frac{L^d}{2\kappa^2 ar^d} \delta\tilde{g}^\prime_{tx} &= \mathcal{Q}\delta A_x - L^d \delta \mathcal{P}_x,  \label{eq33a} \\
\frac{e^2 \mathcal{Q}}{L^{d-2}}\delta \tilde{g}_{tx}^\prime &=  \left(br^{2-d}Z\delta A_x^\prime \right)^\prime + \frac{r^{2-d}Z\omega^2}{b}\delta A_x,    \\
- \frac{k_x\omega \psi_0(\mathbf{k},r)^2}{br^d} \delta\tilde{g}_{tx}&=  \left(\frac{b}{r^d}\psi_0(\mathbf{k},r)^2 \left(\frac{\delta \psi(\mathbf{k},r)}{\psi_0(\mathbf{k},r)}\right)^\prime\right)^\prime + \frac{\omega^2}{br^d} \psi_0(\mathbf{k},r) \delta \psi(\mathbf{k},r),
\end{align}\end{subequations}
where we have defined \begin{equation}
\delta\mathcal{P}_x \equiv \frac{b}{r^d\omega} \int \mathrm{d}^d\mathbf{k} \; k_x |h(\mathbf{k})|^2\psi_0(\mathbf{k},r)^2 \left(\frac{\delta \psi(\omega,\mathbf{k},r)}{\psi_0(\mathbf{k},r)}\right)^\prime.
\end{equation}
In these equations, $\delta A_x$, $\delta \tilde{g}_{tx}$ and $\delta \mathcal{P}_x$ are Fourier components at zero momentum and finite (but small) $\omega$;  they are functions of $r$ only.  The equations of (\ref{emdlin}) are in order:   the $rx$-component of Einstein's equations, the $x$-component of Maxwell's equations, and the $\psi$ wave equation at momentum $\mathbf{k}$;  all of these equations have been integrated over all of space.  Up to the $\mathrm{O}(\omega^2)$ terms in the scalar EOM, the equations close to a finite set of equations for $\delta A_x$, $\delta \mathcal{P}_x$, and $\delta\tilde{g}_{tx}$.    Since we want to work at finite frequencies $\omega \ll T$, but $\omega\tau$ possibly $\gg 1$, it requires some care to argue that we need only consider these three perturbations.

We wrote coefficients of the form $\int \mathrm{d}^d\mathbf{k} \; |h(\mathbf{k})|^2 \psi_0(\mathbf{k},r)^2$ above.   This is analogous to (\ref{eq18}) -- there is a ``missing" $\mdelta$ function.   The reason this arises is that we have integrated over all of space in order to isolate the zero momentum mode $\delta \mathcal{P}_x$ -- this induces an ``infinity" factor analogous to $\mdelta(\mathbf{0})$ in any term which does not vary in space:  for example, the other two terms in (\ref{eq33a}).   Properly regulating this infinity, we recover the equations above -- provided that $|h(\mathbf{k})|^2$ is understood to be ``missing" a factor $\mdelta(\mathbf{0})$, as in Section \ref{sec3}.

Our strategy is to show that the linearized fluctuations need only be computed to $\mathrm{O}(\omega)$ in order to compute the linear response problem to all orders in $\omega\tau$.   This will imply that we can focus on the $\omega\rightarrow 0$ limit of the equations of motion, where we need only solve a linear response problem in three variables, instead of an infinite number.   

Let us begin by studying  the linearized equations of motion when $\omega=0$ exactly -- we will not worry about computing the conductivity just yet.  In this case, the equations of motion reduce exactly to \begin{subequations}\label{eq34}\begin{align}
\frac{L^d}{2\kappa^2 ar^d} \delta\tilde{g}^\prime_{tx} &= \mathcal{Q}\delta A_x - L^d \delta \mathcal{P}_x,  \label{50a} \\
\frac{e^2 \mathcal{Q}}{L^{d-2}}\delta \tilde{g}_{tx}^\prime &=  \left(br^{2-d}Z\delta A_x^\prime \right)^\prime, \label{50b} \\
\delta \mathcal{P}_x^\prime &= -\delta\tilde{g}_{tx}\left[\frac{1}{br^d}\int\mathrm{d}^d\mathbf{k} |h(\mathbf{k})|^2 k_x^2\psi_0(\mathbf{k},r)^2\right],  \label{34c}
\end{align}\end{subequations}
These equations are identical to those of massive gravity \cite{blake1} in the limit $\omega\rightarrow 0$, with the object in brackets in (\ref{34c}) playing the role of the graviton mass.  It is important to remember that it is $\delta \mathcal{P}_x$ which stays O(1) as $\omega\rightarrow 0$, and not $\delta \psi(\omega,\mathbf{k},r)$, which is $\mathrm{O}(\omega)$.   Let us exactly find -- at leading order in $h$ -- all of the solutions of these equations;  counting derivatives we find there must be 4 linearly independent solutions.   The first is spotted by inspection (we will not normalize any of these linearly independent solutions with the correct dimensions): \begin{equation}
\delta A_x = 1, \;\;\; \delta \mathcal{P}_x = \frac{\mathcal{Q}}{L^d}, \;\;\; \delta\tilde{g}_{tx} = 0.
\end{equation}   
In fact, this is the only perturbation that, at leading order in $h$, couples to the $\psi$ sector.    All modes will couple to this sector, but only at $\mathrm{O}(h^2)$, which will prove to be subleading in our computation of the conductivity.   The remaining three modes are, at leading order, modes of the translationally invariant theory with $\psi=0$, and may also be written down exactly.   First there is a ``diffeomorphism" mode \begin{equation}
\delta \tilde{g}_{tx} = 1, \;\;\; \delta A_x = 0, \;\;\; \delta \mathcal{P}_x = 0.  \label{diffmode}
\end{equation}
The next mode may be found by performing a Galilean boost to the static solution: \begin{equation}
\delta A_x =  p + p_{\mathrm{h}}, \;\;\; \delta \tilde{g}_{tx} = 1-ab, \;\;\; \delta\mathcal{P}_x = 0.
\end{equation}where $p_{\mathrm{h}}$ is a constant, and we have conveniently chosen that $\delta\tilde{g}_{tx}(r=0)=0$.   We may fix the value of $p_{\mathrm{h}}$ by requiring that (\ref{50a}) be satisfied at $r=r_{\mathrm{h}}$, and we find, using (\ref{beq}) and (\ref{peq}) (along with the fact that $\mathcal{C}=0$ is fixed by the linearized equations of motion): \begin{equation}
p_{\mathrm{h}} = \frac{2\mpi T L^d}{\kappa^2 r_{\mathrm{h}}^d \mathcal{Q}} = \frac{Ts}{\mathcal{Q}},  \label{galmode}
\end{equation}where $s$ is the entropy density (we have converted between the area of the black hole horizon and the entropy density of the  dual field theory using the Bekenstein-Hawking formula).   We may use the reduction of order method\footnote{Suppose we have a differential equation $f_0y + (f_1y^\prime)^\prime=0$, and $y_0(r)$ is an exact solution of this differential equation.   One can show that a linearly independent solution is $y_1(r)=y_0(r) \int_0^{r} \mathrm{d}s f_1(s)^{-1} y_0(s)^{-2}$.} \cite{tenenbaum} to compute the final solution, by using the fact that both (\ref{galmode}), and the mode we are looking for, solve (at leading order) the differential equation \begin{equation}
\frac{2e^2\kappa^2\mathcal{Q}^2 ar^d}{L^{2d-2}}\delta A_x = \left(br^{2-d}Z\delta A_x^\prime \right)^\prime.
\end{equation}and we find \begin{equation}
\delta A_x = (p(r)+p_{\mathrm{h}})\int\limits_0^r \mathrm{d}s \frac{s^{d-2}}{b(s)Z(s)(p(s)+p_{\mathrm{h}})^2}, \;\;\; \delta \tilde{g}_{tx} = \int\limits_0^r\mathrm{d}s \frac{2\kappa^2 \mathcal{Q} a(s)s^d}{L^d}\delta A_x(s), \;\;\; \delta\mathcal{P}_x = 0.  \label{cgalmode}
\end{equation}
Note that, at leading order in $h$, this mode has a logarithmic divergence at $r=r_{\mathrm{h}}$ in $\delta A_x$ alone.

As $\sigma_{\mathrm{dc}}$ must be finite, and we must have $\delta \tilde{g}_{tx}(r=0)=0$ (we are not sourcing any temperature gradients), at $\omega=0$ the solution to the linearized equations of motion obeying all boundary conditions turns out to be simply \begin{equation}
\delta A_x = \delta A_x^0, \;\;\; \delta \mathcal{P}_x = \frac{\mathcal{Q}}{L^d} \delta A_x^0, \;\;\; \delta\tilde{g}_{tx} = 0.
\end{equation}
We now wish to include corrections when $\omega \ll T, \mu$, but work to all orders in $\omega \tau$.   More precisely, we set all subleading corrections in $\omega/T$ or $\omega/\mu$ to vanish, and then to all orders in $\omega\tau$.   First, we need to discuss the boundary conditions at the black hole horizon.   Both $\delta \mathcal{P}_x$ and $\delta A_x$ obey infalling boundary conditions:  \begin{equation}
\delta A_x, \;   \delta \mathcal{P}_x \sim (r_{\mathrm{h}}-r)^{-\mathrm{i}\omega/4\mpi T}.
\end{equation}
These boundary conditions arise from including the $\mathrm{O}(\omega^2)$ terms in the equations of motion, and demanding that perturbations fall into the black hole horizon (in the boundary theory:  energy is dissipated).    They are, at first glance, non-perturbative in $\omega$, which is frustrating for our purposes.   However, things are not so bad.   Let us fix $r_{\mathrm{h}}-r$ to be arbitrarily small, but send $\omega \rightarrow 0$.   We may then Taylor expand  \begin{equation}
 (r_{\mathrm{h}}-r)^{-\mathrm{i}\omega/4\mpi T} = 1 + \frac{\mathrm{i}\omega}{4\mpi T} \log \frac{r_{\mathrm{h}}}{r_{\mathrm{h}}-r} + \cdots
\end{equation} In a neighborhood of this fixed $r$, the Taylor expansion in $\omega$ must be a solution of the equations of motion, but the $\omega$-dependence of the equations of motions themselves only comes in at $\mathrm{O}(\omega^2)$.  We therefore conclude that the $\mathrm{O}(\omega)$ coefficient must be a solution to the equations of motion evaluated at $\omega=0$.    As we are going to argue later that it is sufficient to only compute $\delta A_x$, $\delta\tilde{g}_{tx}$ and $\delta \mathcal{P}_x$ to $\mathrm{O}(\omega)$, this is a great simplification -- as far as our computation is concerned, (\ref{eq34}) is exact.  Imposing infalling boundary conditions becomes equivalent to imposing the boundary conditions (up to $\mathrm{O}(\omega)$ terms finite at the horizon): \begin{equation}
\delta A_x(r\rightarrow r_{\mathrm{h}}) = \delta A_x^0 \left[1 + \frac{\mathrm{i}\omega}{4\mpi T} \log \frac{r_{\mathrm{h}}}{r_{\mathrm{h}}-r}\right] + \cdots, \;\;\; \delta \mathcal{P}_x(r\rightarrow r_{\mathrm{h}}) = \frac{\mathcal{Q}}{L^d} \delta A_x^0 \left[1 + \frac{\mathrm{i}\omega}{4\mpi T} \log \frac{r_{\mathrm{h}}}{r_{\mathrm{h}}-r}\right] + \cdots.
\end{equation}We impose no specific boundary condition on $\delta \tilde{g}_{tx}$ at $r=r_{\mathrm{h}}$, other than no logarithmic divergences.

Let us now go ahead and compute the $\mathrm{O}(\omega)$ corrections to the linearized modes, at leading order in $h$.   Imposing infalling boundary conditions on $\delta \mathcal{P}_x$,   (\ref{34c}) implies \begin{equation}
\frac{\mathrm{i}\omega}{4\mpi T} \frac{\mathcal{Q}}{L^d} \delta A^0_x \frac{1}{r_{\mathrm{h}}-r} \approx -\frac{\delta \tilde{g}_{tx}}{4\mpi T r_{\mathrm{h}}^d (r_{\mathrm{h}}-r)} \int \mathrm{d}^d\mathbf{k}|h(\mathbf{k})|^2 k_x^2 \psi_0(\mathbf{k},r_{\mathrm{h}})^2.    \label{eq64}
\end{equation}Evidently, this equation is only consistent if $\delta\tilde{g}_{tx}(r=r_{\mathrm{h}}) \sim \omega h^{-2}$ is a finite number.   Since our boundary conditions are that $\delta \tilde{g}_{tx}(r=0)=0$, we conclude that $\delta \tilde{g}_{tx}$ is not a constant at leading order.   (\ref{50b}) then implies that $\delta A_x$ must have a component at $\mathrm{O}(h^{-2})$ as well.   It is consistent in every equation, at $\mathrm{O}(\omega)$, to take $\delta \mathcal{P}_x$ to be $\mathrm{O}(h^0)$, so long as the $\mathrm{O}(\omega h^{-2})$ coefficients in $\delta A_x$ and $\delta \tilde{g}_{tx}$ obey the equations of motion associated with the translationally invariant black hole.   In fact, at $\mathrm{O}(\omega h^{-2})$, we find that the boost mode (\ref{galmode}) is the only mode consistent with all of our boundary conditions:  (\ref{diffmode}) is ruled out by $\delta \tilde{g}_{tx}(r=0)=0$, and (\ref{cgalmode}) by the fact that the logarithmic divergence in $\delta A_x$ occurs at $\mathrm{O}(h^0)$.   Using (\ref{eq64}) we in fact find that the coefficient of the Galilean boost mode in $\delta A_x$ is: \begin{equation}
\delta A_x = \delta A_x^0   - \frac{\mathrm{i}\omega \mathcal{Q}\tau}{\epsilon+P}  (p+p_{\mathrm{h}})  \delta A_x^0  + \mathrm{O}\left(\varepsilon^0\right),
\end{equation}where we have defined \begin{equation}
\frac{\epsilon+P}{\tau} \equiv \frac{L^d}{r_{\mathrm{h}}^d} \int\mathrm{d}^d\mathbf{k} |h(\mathbf{k})|^2 k_x^2\psi_0(\mathbf{k},r_{\mathrm{h}})^2.  \label{taudef2}
\end{equation}
In the next subsection, we show that this $\tau$ is equivalent to that defined via the memory matrix;  for now this is simply a definition of $\tau$.   Note that the logarithmic divergence in $\delta A_x$ is not included, as it is $\mathrm{O}(h^0)$. 

Now, let us explain why we do not have to worry about any further terms, in so far as computing the conductivity at leading order in our limit.   Firstly, let us consider the term that arises from the logarithmic divergence at $\mathrm{O}(h^0)$ in $\delta A_x$.   This will indeed lead to subleading corrections to the Galilean mode described above;  however, these corrections, by dimensional analysis, will scale as $(\omega/T)\mathcal{F}(T/\mu) \delta A_x^0$ for some scaling function $\mathcal{F}$, and are subleading in our limit.   A similar argument holds for the $\mathrm{O}(\omega^2h^{-2})$ term which arises from the fact that when $\omega \tau \gg 1$, the Galilean mode dominates the constant contribution $\delta A_x^0$:   the correction due to infalling boundary conditions on this term $\sim \omega^2 \tau /T$, and this can be neglected in our limit.    Similar arguments will hold for all further corrections at higher orders in $\omega$.   In a nutshell, the linearized equations contain perturbations at most $\sim h^{-2}$, which implies that only the $\omega h^{-2}$ terms need to be included at leading order.   The fact that the boost mode is the only important $\mathrm{O}(\omega)$ contribution to $\delta A_x$ is suggestive of the fact that this is ``hydrodynamic" transport.  

To compute $\sigma(\omega)$ we simply employ (\ref{peq2}) to find the near boundary behavior of $p(r)$.   We thus find \begin{subequations}\begin{align}
\delta A_x^{(0)}(\omega) &= \delta A_x^0 \left[1 - \frac{\mathrm{i}\omega\mathcal{Q}\tau}{\epsilon+P}\left( \mu + \frac{Ts}{\mathcal{Q}}\right)\right]= \delta A_x^0 \left[1 - \frac{\mathrm{i}\omega\tau (\mu \mathcal{Q}+Ts)}{\epsilon+P}\right] =  \delta A_x^0(1-\mathrm{i}\omega \tau),   \label{deltaax0} \\
\delta A_x^{(1)}(\omega) &= -\frac{e^2\mathcal{Q}}{L^{d-2}} \left(- \frac{\mathrm{i}\omega\mathcal{Q}\tau}{\epsilon+P}\right)\delta A_x^0,
\end{align}\end{subequations}where we have used the thermodynamic identity \begin{equation}
\epsilon+P = \mu\mathcal{Q}+Ts
\end{equation} in the last step of (\ref{deltaax0}).   Recall that $\delta A_x^{(1)}/\delta A_x^{(0)}$ is related to $\sigma(\omega)$ by (\ref{sigmaomega}).  It is straightforward from here to obtain (\ref{drude}),  $\sigma(\omega) = \sigma_{\mathrm{dc}}/(1-\mathrm{i}\omega \tau)$, and the expression (\ref{sigmadc}) for $\sigma_{\mathrm{dc}}$ in the limit where $\omega/T\rightarrow 0$ first, and $\omega \tau$ is held finite.

That Drude physics arises is not particularly surprising for a few reasons.   Intuitively, we expect via the fluid-gravity correspondence \cite{minwalla} that a holographic system, weakly perturbed by translational symmetry breaking, behaves in the same way as a fluid with momentum relaxation would:  see also \cite{herzog}.    Additionally, it was shown via matched asymptotic expansions in \cite{davison} that certain massive gravity theories contain Drude peaks in the limit of weak graviton mass, at $T=0$.   Due to the equivalence between the holographic set-ups above and massive gravity as $\omega\rightarrow 0$ \cite{blake2} (with the caveat that the graviton mass becomes a function of $r$), this result makes sense.   It is pleasing nonetheless to see the Drude peak emerge from a direct calculation for a wide range of holographic theories at finite $T$.

\subsection{Equivalence with the Memory Function Approach}
Our only remaining task is to show that the $\tau$ defined above is equivalent to the $\tau$ defined in (\ref{taudef}), defined in terms of the leading order imaginary behavior of $G^{\mathrm{R}}_{\mathcal{OO}}(\omega\rightarrow 0)$.   By the holographic dictionary for retarded Green's functions, we impose infalling boundary conditions at the black hole horizon $r=r_{\mathrm{h}}$, as before.  Near the black hole horizon: \begin{equation}
\psi(\mathbf{k},r,\omega\rightarrow 0)\approx  \psi(\mathbf{k},r,\omega=0)\left[1+ \frac{\mathrm{i}\omega}{4\mpi T} \log \frac{r_{\mathrm{h}}}{r_{\mathrm{h}}-r}\right].   \label{ch}
\end{equation}
As before, at $\mathrm{O}(\omega)$, $\psi$ solves its $\omega=0$ equation of motion.   Also note $\psi(\mathbf{k},r,\omega=0) = \psi_0(\mathbf{k},r)$, with $\psi_0$ the finite scalar profile we defined previously.  We wish to ``propagate" this small imaginary piece from the horizon to the AdS boundary.   This directly allows us to compute $\mathrm{Im}(G^{\mathrm{R}}(\omega))/\omega$ by using the AdS/CFT dictionary at $r=0$.    We compute the linearly independent solution to $\psi$'s equation of motion with the reduction of order technique: \begin{equation}
\psi_1(\mathbf{k},r) \equiv \frac{\psi_0(\mathbf{k},r)}{L^d} \int\limits_0^r \mathrm{d}s \frac{s^d}{b(s) \psi_0(\mathbf{k},s)^2}
\end{equation}We assume that $\psi_0$ has the same asymptotics as before:  $\psi_0(\mathbf{k},r\rightarrow 0) \approx L^{-d/2}r^{d+1-\Delta}$, and have normalized $\psi_1$ conveniently.  The integral as written above is convergent so long as $\Delta > (d+1)/2$, and we have assumed this property previously.  In fact, this solution also has the $r\rightarrow 0$ asymptotics we wish: as $b\approx 1$ near the AdS boundary, we find that  \begin{equation}
\psi_1(\mathbf{k},r)\approx L^{-d/2} \frac{r^\Delta }{2\Delta-d-1}, \;\;\; (r\rightarrow 0).
\end{equation}
Near the black hole horizon, $\psi_1$ is divergent -- this divergence implies that the imaginary contribution to $\psi$ is (for the purposes of our calculation) proportional to $\psi_1$.   As this divergence is associated with near-horizon physics, its coefficient is completely independent of the UV or intermediate scales in the geometry:   \begin{equation}
\psi_1(\mathbf{k},r) \approx \frac{r_{\mathrm{h}}^d}{4\mpi T\psi_0(\mathbf{k},r_{\mathrm{h}}) L^d}\log \frac{r_{\mathrm{h}}}{r_{\mathrm{h}}-r}+ \text{finite}, \;\;\; (r\rightarrow r_{\mathrm{h}}).   \label{psi1}
\end{equation}Comparing (\ref{ch}) and (\ref{psi1}) we conclude that to order $\omega$, \begin{equation}
\psi(\mathbf{k},r,\omega) \approx \psi_0(\mathbf{k},r) + \mathrm{i}\omega \frac{L^d \psi_0(\mathbf{k},r_{\mathrm{h}})^2}{r_{\mathrm{h}}^d} \psi_1(\mathbf{k},r) + \mathrm{O}\left(\omega^2\right),  \label{fineq}
\end{equation}
The function $\psi_0$ is real.  Therefore using (\ref{expo}), it is easy to conclude that \begin{equation}
\lim_{\omega\rightarrow 0} \frac{\mathrm{Im}\left(G^{\mathrm{R}}_{\mathcal{OO}}(\mathbf{k},\omega)\right)}{\omega} = \frac{L^d \psi_0(\mathbf{k},r_{\mathrm{h}})^2}{r_{\mathrm{h}}^d}   \label{imgr}
\end{equation}    
Comparing (\ref{taudef}), (\ref{taudef2}) and (\ref{imgr}), we see that the conductivity of this metal, computed directly via holography, is identical to the result computed with the memory function formalism, without using any holography.   These results are valid in the limit when $\omega,\tau^{-1}$ are vanishingly small compared to $T$ and $\mu$, the regime of validity where the approximation (\ref{sigmamm}) holds.  

\subsection{Seebeck Coefficient}
Let us briefly discuss thermoelectric transport coefficients \cite{thermoel1, thermoel2, thermoel3, thermoel5, thermoel4}.   One such coefficient is immediately computable given our discussion of the conductivity:  the Seebeck coefficient $\alpha$, defined by \begin{equation}
q_x = \alpha T E_x,
\end{equation}where $q_x$ is the heat flow density in the $x$ direction.   Using \begin{equation}
q_x = \langle T^{tx}\rangle - \mu \langle J^x\rangle,
\end{equation}and reading off the expectation value of $\langle T^{tx}\rangle$ in the linearized modes computed in this section, it is straightforward to recover \begin{equation}
\alpha(\omega) = \frac{s\mathcal{Q}\tau}{\epsilon+P} \frac{1}{1-\mathrm{i}\omega\tau}, 
\end{equation}in agreement with hydrodynamics \cite{hkms} and the memory function formalism (use that the heat-momentum susceptibility $\chi_{QP} = sT$).     We also expect that a calculation of the thermal conductance $\bar\kappa$, defined by \begin{equation}
q_x = \left. -\bar \kappa \partial_x T\right|_{E_x=0},
\end{equation}would find, in agreement with hydrodynamics and the memory function formalism, \begin{equation}
\bar\kappa(\omega) = \frac{s^2 T \tau}{\epsilon+P} \frac{1}{1-\mathrm{i}\omega\tau}.
\end{equation}Computations of $\bar\kappa(\omega)$ are a bit more involved \cite{thermoel1, thermoel2, thermoel3, thermoel5, thermoel4} and we will not pursue them further.

\section{Conclusion}
In this work, we have demonstrated an exact correspondence between the memory function approach and a holographic approach to transport in a strongly correlated ``strange" metallic phase of matter without quasiparticles, in the limit of finite density and slow momentum relaxation.      This result crispens the qualitative scaling agreements noted in special cases earlier \cite{blake2, lss}.   We have also pointed out the emergence of a universal Drude peak in these holographic models.   These results are to be expected  on physical grounds, but it is nonetheless instructive to see them explicitly verified.

The reason that we add a fourth scalar field $\psi$, instead of say adding perturbations to the dilaton, is that the EMD background is only corrected at $\mathrm{O}(h^2)$;   it would receive corrections at $\mathrm{O}(h)$ if the dilaton had a component which breaks translational symmetry.   As in \cite{blake2, lss}, adding a new scalar simplifies the resulting computation of $\sigma(\omega)$.  It would be worthwhile to explicitly show that the correspondence with the memory matrix survives even if we break translational symmetry through one of the EMD fields:  for example, using results of \cite{donos1409}, it should be possible to do this for spatially-dependent chemical potentials.   

One of the advantages to a holographic computation is that we are formally not restricted to the limit of weak momentum relaxation.    Approaches including massive gravity and Q-lattices, which produce translation invariant geometries, allow for analytic control in this limit.   Perhaps techniques similar to ours can remain valid even in this limit of strong momentum dissipation.  While it is now clear that such approaches mimic more conventional ``lattices" or disorder when momentum relaxation is weak, it is less clear whether this analogy remains when momentum relaxation is strong.   It is possible that more exotic phases of holographic matter, analogous to a quantum glass or many-body localized phase \cite{huse}, will arise when translational symmetry is explicitly broken strongly.  Alternately, efficient momentum relaxation without ``glassy" physics may be responsible for scaling properties of cuprate strange metals \cite{hartnollscale2};  perhaps massive gravity/Q-lattice approaches are powerful tools for these problems.   More work  in this direction is warranted.

\addcontentsline{toc}{section}{Acknowledgements}
\section*{Acknowledgements}
I would like to thank Mike Blake, Blaise Gout\'eraux, Sean Hartnoll, Subir Sachdev and Koenraad Schalm for helpful comments on the manuscript.    This research was supported by a teaching fellowship at Harvard University, the NSF under Grant DMR-1360789, and MURI grant W911NF-14-1-0003 from ARO.

\end{document}